\documentclass[twocolumn,superscriptaddress,pre,letterpaper]{revtex4}
\usepackage{hyperref}
\usepackage{xurl} 
\usepackage{amsmath}
\usepackage{amsbsy}
\usepackage{mathtools}
\usepackage{bbm}
\usepackage{dcolumn}
\usepackage{enumitem}
\usepackage{graphicx}
\usepackage{float}
\usepackage{xcolor}
\usepackage{multirow}
\usepackage{amssymb}

\hypersetup{
    colorlinks=true,
    citecolor=green!60!black,
    linkcolor=blue,
    urlcolor=blue,
}

\newcommand\bm{\boldsymbol}
\DeclareMathOperator*{\argmax}{arg\,max}

\begin{document}

\title{Model inference for ranking from pairwise comparisons}
\author{Daniel Sánchez Catalina}
\affiliation{Department of Engineering, University of Cambridge, CB2 1PZ, United Kingdom}
\author{George T. Cantwell}
\affiliation{Department of Engineering, University of Cambridge, CB2 1PZ, United Kingdom}

\begin{abstract}
We consider the problem of ranking objects from noisy pairwise comparisons, for example, ranking tennis players from the outcomes of matches. We follow a standard approach to this problem and assume that each object has an unobserved strength and that the outcome of each comparison depends probabilistically on the strengths of the comparands. However, we do not assume to know a priori how skills affect outcomes.
Instead, we present an efficient algorithm for simultaneously inferring both the unobserved strengths and the function that maps strengths to probabilities.
Despite this problem being under-constrained, we present experimental evidence that the conclusions of our Bayesian approach are robust to different model specifications.
We include several case studies to exemplify the method on real-world data sets.
\end{abstract}

\maketitle


People habitually rank things.
Sometimes ranking is purely for fun---e.g., ranking your favorite movies---sometimes it is a matter of life and death---e.g., ranking candidate organ recipients---and often it will be in-between these extremes.
Numerous algorithms have been developed for ranking.

Motivated by chess, Zermelo formalized the problem of ranking from comparisons \cite{zermelo_berechnung_1929, Glickman2013}.
Zermelo's work, along with the work of Bradley and Terry \cite{bradley_rank_1952}, Ford \cite{ford_solution_1957} Thurstone \cite{thurstone_law_1927}, Mosteller \cite{mosteller_remarks_1951}, Kendall \cite{kendall_further_1955}, and Elo \cite{elo_rating_1986, aldous_elo_2017} are foundational in the ranking literature; see Ref.~\cite{davidson_bibliography_1976} for a bibliography of early work and Refs.~\cite{ali_minimum_1986,  debacco_physical_2018, negahban_ranking_2012, cantwell_belief_2022, jerdee_luck_2023, trueskill} for a selection of more recent works.
We consider the same basic setting.

We seek to rank a set of $n$ things according to their strength or ability, but where direct measurement of ability is not possible.
Instead, we observe the outcomes of stochastic comparisons, e.g., for tennis we might observe ``player~$i$ beat player~$j$''.
The standard approaches assume each item has a latent strength parameter, attempt to infer these parameters, and then rank items using the inferred strengths.

While Zermelo (and many subsequent studies) used the method of maximum likelihood,
accurate and precise estimation of parameters is intrinsically challenging due to sparse, noisy, and seemingly contradictory data.
Bayesian methods have been developed to alleviate this \cite{davidson_bayesian_1973, Leonard_alt_bayes_1977, ADAMS20051191, SHEV20121523, caron_efficient_2012}.
Considerable attention has also been given to generalization of the setting.
For example, the setting has been generalized to include draws, to assign credit in team sports, to allow for multiple dimensions of strength, or to include covariates \cite{davidson1969multivariate, davidson_extending_1970, Dittrich_modelling_1998, joachims_optimizing_2002, minka2018trueskill, caron_efficient_2012, hunter_MM_2004, NIPS2004_825f9cd5, santi2025bradleyterrystochasticblockmodel}.

In contrast, we look at the understudied but important problem of inferring the \emph{model}.
Our contribution is an efficient algorithm to jointly infer both the model and the strength parameters in the pairwise comparison setting.
The key distinction is that, in addition to the strengths being unknown, we assume that the function that maps strengths to win-loss probabilities is also unknown.

We implement both Chebyshev-based and neural network-based algorithms to find this unknown function.
We find consistency with ground-truth when applying both algorithms to synthetic data, and consistency between the algorithms when applied to real data sets.
We also look at men's professional tennis where we find improved representations of uncertainty and improved predictive accuracy.
Notably, we are able to overcome a bookmaker's profit margin using only the win-loss records as input data.

\section{The pairwise comparison setting}
To rank $n$ objects our only input data will be the $n \times n$ matrix $w$ with entries
\begin{equation}
  w_{ij} = \text{number of times }i \text{ beat } j.
\end{equation}
In general $w$ will not be symmetric and can include a large number of $0$s.
For linguistic ease, we will refer to the objects as \emph{players} and the comparisons as \emph{matches}, although mathematically this changes nothing.

We assume there is a single numerical quantity that can be assigned to each player to represent their strength or skill.
We denote by $x_i$ the skill of player~$i$ and assume that conditional on these skills the matches are independent. 
The likelihood is 
\begin{equation}
\begin{aligned}
	P_b(w \vert \bm{x}) &=  \prod_{i<j} \binom{w_{ij} + w_{ji}}{w_{ij}} b(x_i, x_j)^{w_{ij}} b(x_j, x_i)^{w_{ji}}  \\
    &=\prod_{i,j} \sqrt{\tbinom{w_{ij} + w_{ji}}{w_{ij}}} \, b(x_i, x_j)^{w_{ij}}
\end{aligned}
  \label{eq:ranking_likelihood}
\end{equation}
where $b(x, y)$ is the probability that a player with skill $x$ beats a player with skill $y$. 
Alternatively, if we know the order in which matches occur the likelihood is
\begin{equation}
  P_b(w \vert \bm{x}) = \prod_{i, j} b(x_i, x_j)^{w_{ij}}.
\end{equation}
(In subsequent analysis the binomial terms will drop out and the distinction between these two cases is moot.)

To complete the model specification one must make a choice for the function $b(x,y)$.
The assumptions of Zermelo \cite{zermelo_berechnung_1929} and of Bradley and Terry \cite{bradley_rank_1952} are equivalent to $b(x,y)$ being the logistic function, 
\begin{equation}
  b(x,y) = \frac{1}{1 + e^{y-x}}
\end{equation}
and this is referred to as the Bradley-Terry model.
This choice leads to a log-concave likelihood and hence it is straightforward to numerically estimate the skills from the match results.
In an essentially equivalent approach, we may chose $b(x,y) = \Phi(x-y)$, where $\Phi$ is the cumulative distribution function of the standard normal distribution \cite{thurstone_law_1927, mosteller_remarks_1951}.

Of course, making the wrong choice for $b(x,y)$ would lead to incorrect inferences about the skills $\boldsymbol{x}$, and potentially by a large margin.
This problem has been less widely addressed though it has been long noted (e.g., it is discussed by Davidson and Solomon~\cite{davidson_bayesian_1973} and Keener~\cite{keener_perron_1993}).
A simple parametric approach would add free parameters to $b(x,y)$.
For example, one approach assumes $b(x, y) = \alpha/2 + {(1 - \alpha)}/(1 + e^{\beta(y - x)})$, and the parameters $\alpha \in [0, 1]$ and $\beta > 0$ are fit simultaneously to the skills \cite{jerdee_luck_2023}.

Instead of assuming any particular functional form for $b(x,y)$ we will represent it using either Chebyshev or neural network approximants.
We will fit the model using an expectation-maximization (EM) algorithm, following a very similar procedure to Newman and Peixoto \cite{NewmanPeixoto2015}, where an EM algorithm was developed to study community structure in networks.
To this end, we first proceed on the assumption that $b(x,y)$ is already known.

\subsection{Bayesian ranking when $b(x,y)$ is known}
Even when $b(x,y)$ is known it may not be possible to reliably estimate $\bm{x}$.
If we were to observe an increasing number of matches between a fixed set of players, then consistent estimation of $\bm{x}$ should be possible if $b(x,y)$ were known. 
In fact, even if the number of players is increased then so long as the number of matches per player also increased it may be possible to accurately infer parameters \cite{simons_asymptotics_1999}.
However, in the real-world the number of matches often cannot grow faster than linearly in the number of players.
For example, human lives are finite and this fact places an upper bound on the number of tennis matches any individual could play, leaving fundamental uncertainty about $\bm{x}$.
To see this more formally, note that the Fisher information is
\begin{equation}
	-E\bigg[ \frac{\partial^2 \log P(w \vert \bm{x})}{\partial x_i^2} \bigg] 
	= k \bigg(\frac{1}{n} \sum_j \frac{b'(x_i,x_j)^2}{b(x_i,x_j)b(x_j,x_i)} \bigg)
\end{equation}
where $k$ is the expected number of matches played by individual~$i$ and $b'$ is the derivative of $b$ with respect to the first argument.
Even if all other parameters $x_j$ were known, unbiased estimators for $x_i$ will have variance proportional to $1/k$ and so in the sparse (and realistic) regime, estimation of $x_i$ carries intrinsic uncertainty.

For this reason even if the true function $b(x,y)$ were known we should advocate a Bayesian approach, i.e., placing a prior on the skills and considering their posterior distribution.

We propose using a uniform prior for $\bm{x}$ in $\left[0,1\right]^n$ as this is a ``natural'' representation for ranking.
First, it is only reasonable to assume that all $x_i$ are independent and identically distributed in the prior.
In this case any choice of continuous distribution is equivalent up to a change of variables and a corresponding change to $b(x,y)$.
Second, the uniform prior has the unique interpretation as percentiles. For example, a player with $x_i = 0.7$, would be a $70$th percentile player.
Hence, we consider
\begin{equation}
  P_b(\bm{x} \vert w) = \frac{\prod_{i,j} b(x_i, x_j)^{w_{ij}}}{\int \prod_{i,j} b(u_i, u_j)^{w_{ij}} \mathrm{d} \bm{u}}.
  \label{eq:skill_posterior}
\end{equation}

As is typical for Bayesian approaches, the distribution in Eq.~\eqref{eq:skill_posterior} is not easy to evaluate,
but we use a fast and accurate approximation.
We follow the approach of Cantwell and Moore~\cite{cantwell_belief_2022} which combines belief propagation and Chebyshev approximants to efficiently estimate the posterior.
By standard arguments \cite{mezard_information_2009, mackay_information_2003, Moore2017TheCS} we define a message function from player $j$ to $i$ as
\begin{equation}
	\mu_{i \leftarrow j}(x) \propto \prod_{k (\neq i)} \int \mu_{j \leftarrow k}(y)
	b(x, y)^{w_{jk}} b(y, x)^{w_{kj}}
	\mathrm{d}y
	\label{eq:message}
\end{equation}
where normalization is fixed so that $\int \mu_{i \leftarrow j}(x) \mathrm{d}x = 1$.
By replacing functions with Chebyshev approximants the above integral becomes a matrix multiplication.
All messages functions can then be found by a simple iteration scheme (see Ref.~\cite{cantwell_belief_2022} for further details).

The messages themselves are not of direct interest, but from them we can approximate marginal distributions.
For example, the posterior marginal distribution for the skill of player~$i$ is well approximated by
\begin{equation}
	\mu_{i}(x) \propto \prod_{j (\neq i)} \int \mu_{i \leftarrow j}(x_j)
	b(x, x_j)^{w_{ij}} b(x_j, x)^{w_{ji}}
	\mathrm{d}x_j
\end{equation}
while the joint marginal distribution for the skill of player~$i$ and $j$ is well approximated by
\begin{equation}
	\mu_{ij}(x, y) \propto  \mu_{j \leftarrow i}(x) \mu_{i \leftarrow j}(y)
	b(x, y)^{w_{ij}} b(y,x)^{w_{ji}}.
	\label{eq:joint_bp_marginal}
\end{equation}
The ability to efficiently approximate the joint marginal of $x_i$ and $x_j$ using Chebyshev approximants and belief propagation will be enormously useful for estimating $b(x,y)$, as we presently see.

\subsection{Inferring the kernel $b(x,y)$}
To pick among different choices for $b(x,y)$, we consider the \emph{model evidence}
\begin{equation}
  P_b(w) = \int \prod_{i,j} b(x_i, x_j)^{w_{ij}} \mathrm{d}\bm{x}.
  \label{eq:model_evidence}
\end{equation}
Of course, without further restriction the function $b$ is not identifiable.
To see this, let $\pi$ be any measure preserving transformation and define $b^{\star}(u, v) = \hat{b}(\pi(u), \pi(v))$.
Then
\begin{equation}
\begin{aligned}
	P_{b^{\star}}(w) &= \int \prod_{i,j} b^{\star}(x_i^{\star}, x_j^{\star})^{w_{ij}} \mathrm{d}\bm{x}^\star \\
	 &= \int \prod_{i,j} \hat{b}(\pi(x_i^{\star}), \pi(x_j^{\star}))^{w_{ij}} \mathrm{d}\bm{x}^\star = P_{\hat{b}}(w)
\end{aligned}
\end{equation}
where the last equality holds by the change of variables $x_i = \pi(x_i^{\star})$,
and hence for every $\hat{b}$ there are at least as many equivalent $b^\star$ as there are measure preserving transformations.

This non-identifiability is a generic problem when inferring functions and the solution is to make strong assumptions about the space of acceptable functions.
We hence find $b$ by maximizing
\begin{equation}
	\log P_b(w) + R\left[ b \right]
	\label{eq:objective}
\end{equation}
where $R\left[ b \right]$ is the penalty that encodes our assumptions about $b(x,y)$.
Equivalently, we can interpret $e^{R\left[ b \right]}$ as a (non-normalized) prior probability for function $b(x,y)$.

We consider two separate and conceptually different approaches:
(i) a Chebyshev prior that places derivative constraints on $b(x,y)$ and (ii) parameterization of $b$ via a neural network.
We find good agreement between both approaches which is evidence that the approach is robust to poor specification of $b(x,y)$.

To optimize Eq.\eqref{eq:objective} and find $b(x,y)$, first note that by Jensen's inequality for any distribution $Q(\bm{x})$ we have
\begin{equation}
  \log P_b(w) \geq \int Q(\bm{x}) \log \bigg(
  \frac{\prod_{i,j} b(x_i, x_j)^{w_{ij}}}{Q(\bm{x})} \bigg) \mathrm{d}\bm{x}.
  \label{eq:jensen}
\end{equation}
Setting $Q(\bm{x}) \propto \prod_{i,j} b(x_i, x_j)^{w_{ij}}$ saturates the inequality and hence double maximization of the right hand side of Eq.~\eqref{eq:jensen} with respect to both $Q$ and $b$ is equivalent to maximization of the left with respect to $b$.
To maximize the right hand side of Eq.~\eqref{eq:jensen} with respect to $b$ for fixed $Q$ we would maximize
\begin{equation}
 \sum_{i,j} w_{ij}  \iint Q_{ij}(x,y) \log b(x,y) \mathrm{d}x \mathrm{d}y
\end{equation}
where $Q_{ij}(x, y)$ is the marginal distribution for the skill of $i$ and $j$ in $Q(\bm{x})$.

This naturally leads to the following iterative algorithm to optimize Eq.~\eqref{eq:objective}.
First, make an initial guess for $b(x, y)$. Then, iteratively refine the estimate by:
\begin{enumerate}
	\item Computing the marginal distributions from Eqs.~\eqref{eq:message} and \eqref{eq:joint_bp_marginal} (i.e. belief propagation) and setting
    \begin{equation}
      \overline{Q}(x, y) = \sum_{i,j} w_{ij} \mu_{ij}(x, y).
    \end{equation}
  \item Updating the estimate of $b(x,y)$ by setting 
\begin{equation}
  b = \argmax_{b} \bigg\{ \iint \overline{Q}(u, v) \log b(u,v) \mathrm{d}u \mathrm{d}v + R\left[ b \right]\bigg\}.
  \label{eq:M_step}
\end{equation}
\end{enumerate}

If the functions $\mu_{ij}(x, y)$ from Eq.~\eqref{eq:joint_bp_marginal} were exact representations of the marginal distributions, and if the optimization in Eq.~\eqref{eq:M_step} were exact, then this algorithm would converge to a (local) maximum of our objective function, Eq.~\eqref{eq:objective}.
In our approach most steps are approximate but we will later demonstrate good performance despite this.

\subsection{Two alternative priors for $b(x,y)$}
We have a two important constraints for $b(x, y)$ that must be respected.
First, $b(x,y)$ must be a valid probability so its codomain must be $\left[0, 1\right]$.
For convenience we can re-parameterize to $f(x,y)$ with
\begin{equation}
  b(x, y) = \frac{1}{1 + e^{-f(x,y) } }
\end{equation}
and where $f(x,y)$ can take any real value.
Second, because the probability that $i$ beats $j$ or $j$ beats $i$ must be $1$ we have $b(x,y) + b(y,x) = 1$ and hence the anti-symmetry constraint
\begin{equation}
f(x,y) = -f(y,x).
\end{equation}
Otherwise we are left with considerable freedom for parameterizing $f(x,y)$.
We consider two different methods.

{\bf{Chebyshev prior.}}
We can represent the function $f(x,y)$ by a Chebyshev expansion
\begin{equation}
  f(x,y) = \sum_{\alpha, \beta} c_{\alpha \beta} T_{\alpha}(2x-1) T_{\beta}(2y-1)
\end{equation}
where $T_k$ is the $k$th Chebyshev polynomial.
To place a prior on $f$ we make two assumptions.

First, we assume $f(x, y)$ should be reasonably smooth.
To this end we penalize the coefficients $c_{\alpha \beta}$ for large $\alpha$ and $\beta$ according to
\begin{equation}
	R\left[ f \right] =  -\frac{1}{64} \sum_{\alpha=0}^{L-1}\sum_{\beta=0}^{L-1} \left(\left(\alpha^2 + \beta^2\right) c_{\alpha \beta} \right)^2
\end{equation}
with the hard limit that $c_{\alpha,\beta}=0$ when $\alpha \geq L$ or $\beta \geq L$.
We assume an upper cut-off of $L=32$, though this should not be too important because the quadratic regularization harshly penalizes higher-order coefficients.

Additionally, we enforce that $f(x, y)$ is monotonic in both arguments.
To achieve this, we first note that, since $f(x, y)$ a degree $L=32$ Chebyshev polynomial, it is entirely determined by its values at the Chebyshev nodes $f(x_k, x_l)$ where
\begin{equation}
x_k = \frac{1}{2} - \frac{1}{2} \cos \Big( \frac{(k+\frac{1}{2})\pi}{L} \Big).
\end{equation}
To enforce both symmetry and monotonicity, we represent $f(x_k, x_m)$ at the Chebyshev nodes by
\begin{equation}
f(x_k, x_m) = \Big( \sum_{i=k}^{m} \sum_{j=i+1}^{m} \vert g_{ij} \vert^p \Big)^{1/p}
- \Big( \sum_{i=m}^{k} \sum_{j=i+1}^{k} \vert g_{ij} \vert^p \Big)^{1/p}
\end{equation}
where $g_{ij}$ is now an entirely unconstrained $L \times L$ upper-triangular matrix and we arbitrarily set $p=8$.

With this representation, we optimize the objective
\begin{equation}
\begin{aligned}
	\iint \overline{Q}(u, v) \log \Big( &\frac{1}{1 + e^{-f(u,v)}} \Big) \mathrm{d}u \mathrm{d}v  \\
    &-\frac{1}{64} \sum_{\alpha,\beta} \left(\left(\alpha^2 + \beta^2\right) c_{\alpha \beta} \right)^2 
\end{aligned}
\end{equation}
with respect to all $c_{\alpha\beta}$ using Newton's method.
The two-dimensional integral is computed by Clenshaw–Curtis quadrature and derivatives are computed by automatic differentiation.

{\bf{Neural network prior.}}
Alternatively, we can parameterize $f(x,y)$ as a neural network $g_\theta(x,y)$, with trainable weights $\theta$.
The architecture we use is a fully connected feed-forward network (a multilayer perceptron) with two hidden layers of width $64$ and ReLU activation functions.
To respect the symmetry constraint we output $f(x,y) = g_\theta(x,y) - g_\theta(y, x)$.

To train the neural network $g_\theta(x,y)$ so that it (approximately) maximizes Eq.~\eqref{eq:objective}, we 
sample a large number of pairs $(x_s, y_s)$ proportional to $\overline{Q}(x, y)$.
These samples form a training set through the loss function 
\begin{equation}
-\sum_s \log \Big( \frac{1}{ 1 + e^{g_\theta(y_s, x_s) - g_\theta(x_s - y_s)}} \Big) + \frac{\left\vert \theta \right\vert^2}{\sum_{i,j}w_{ij}} 
\end{equation}
where $\left\vert \theta \right\vert^2 = \sum_\alpha \theta_\alpha^2$ is the quadratic norm of the parameters of the neural network. The loss is minimized with the Adam optimizer using the default settings in Pytorch.
Note, we scale the relative strength of the two terms by the number of observed matches so that the algorithm has a Bayesian interpretation, namely a Gaussian prior on the parameters of the neural network.

\section{Results}

\subsection{Consistency}

\begin{figure*}
    \centering
    \includegraphics[width=0.75\linewidth]{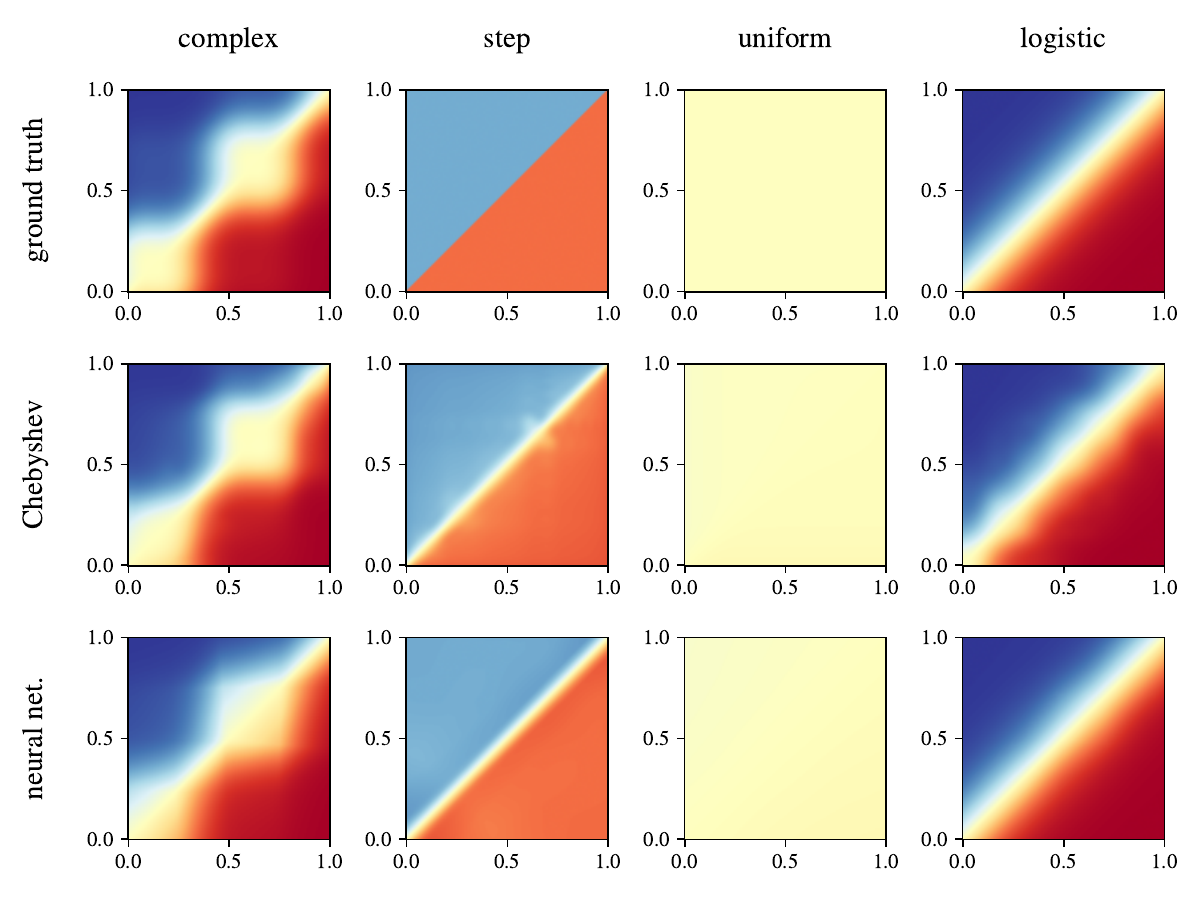}
    \caption{Synthetic data experiments. 
    Matches were simulated between $1024$ players, each of whom took part in $64$ matches.
    Skills were assigned uniformly at random and outcome data   was generated using $4$ different kernels: complex, step, uniform, and logistic.
    On the top row we show the ground-truth kernel, $b(x,y)$.
    On the middle row, we show the kernel inferred from the win-loss-record by the Chebyshev method.
    On the bottom row, we show the kernel inferred from the win-loss-record by the neural network method.
    }
    \label{fig:synthetic_comparison}
\end{figure*}

As an initial test to ensure that both methods converge to similar solutions, we simulate competitions between $1024$ players, each of which takes part in $64$ matches.
Ground-truth skills are assigned uniformly from $0$ to $1$, and we experiment with $4$ different kernel functions $b(x,y)$.
Figure~\ref{fig:synthetic_comparison} shows the ground-truth kernels, and the inferred kernel using both the Chebyshev method and the neural network.
Good agreement is found between both methods and the ground truth.

\begin{figure*}
    \centering
    \includegraphics[width=1\linewidth]{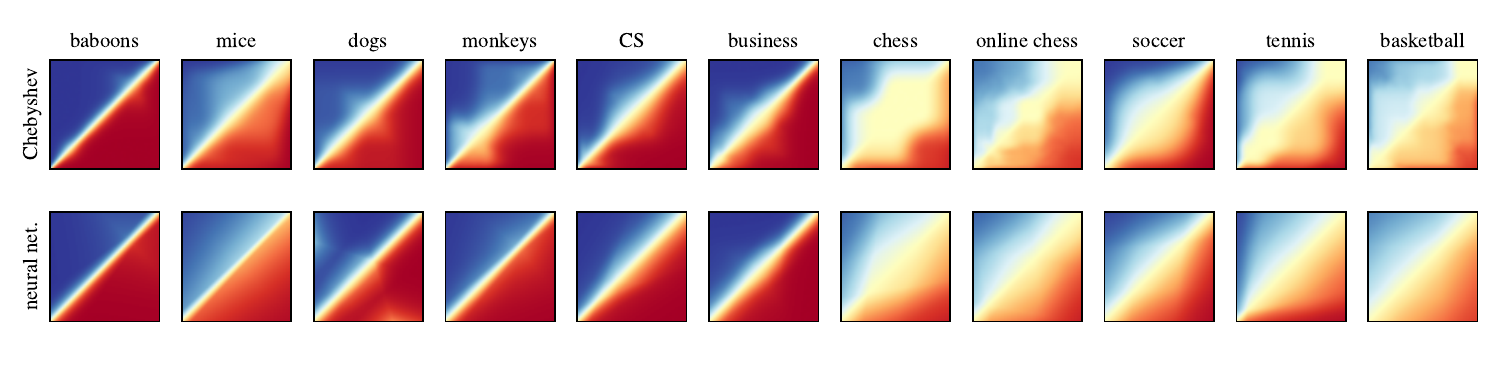}
    \caption{Inferred kernels $b(x,y)$ from 11 real-world datasets, described in Table~\ref{tab:dominance_dataset_descriptions}.
    On the top row we show the inferred kernels from the win-loss-record using the Chebyshev method while the bottom shows the neural network.
    }
    \label{fig:real_data_kernels}
\end{figure*}

Next we explore $11$ real-world data sets.
The data are informative on different hierarchies including professional and amateur sports, academic prestige, and animal dominance (see Table \ref{tab:dominance_dataset_descriptions} for descriptions of all datasets). 
Kernel fits are shown in Fig.~\ref{fig:real_data_kernels}.

A split is visually apparent.
The first $4$ datasets are animal dominance interactions; all $4$ look similar to one another.
Likewise the final $5$ are human games and look similar to each other.
Interestingly the middle $2$, which correspond to academic hiring, look more similar to animal dominance than human games.

Clearly we cannot access the ``ground-truth'' for these data---it is not clear such a thing exists since the model is presumably misspecified.
Nevertheless, we observe a strong agreement between the kernels inferred by both methods. For example the optimal matching between the Chebyshev and neural network functions is the identity map.
The consistency between our very different specifications for $b(x,y)$ is an indication that we are finding true signal in the data.

\subsection{Case study: Association of Tennis Professionals (ATP)}

We now proceed with a more in depth case study of men's professional tennis.

First, we compare our ranking method to the official ATP end-of-year rankings for $2024$.
Using our methods we assign an individual percentile to each player by computing their mean skill in the posterior distribution, i.e.,
\begin{equation}
r_i = 100 \int x \mu_i(x) \mathrm{d}x.
\end{equation}
In contrast, the ATP ranks players using a points-based system, where victory in a match confers a predetermined number of points depending on the tournament round and level.

Ranking by either ATP points or by model inferred percentile shows Jannik Sinner as the top player of the year.
However, his ATP points show him as a very large outlier whereas the gap between inferred percentiles are more moderate.
The ordering also changes slightly. 
For example, in part due to injury, Novak Djokovic won fewer ATP points than Taylor Fritz but his inferred skill was higher.

\begin{table}
\centering
\small
\setlength{\tabcolsep}{4pt} 
\renewcommand{\arraystretch}{1.1} 
\begin{tabular}{clccc}
\hline
\textbf{rank} & \textbf{name} & \textbf{country} & \textbf{ATP} & $\boldsymbol{r}$ \\
\hline
1  & Jannik Sinner    & \raisebox{-0.3\height}{\includegraphics[width=0.9em]{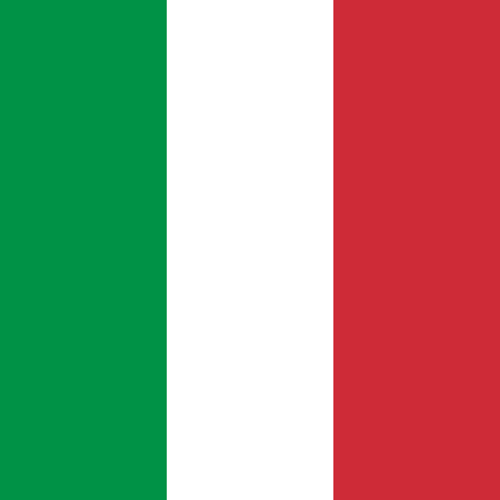}} & 11,830 & 99.7 \\
2  & Carlos Alcaraz   & \raisebox{-0.3\height}{\includegraphics[width=0.9em]{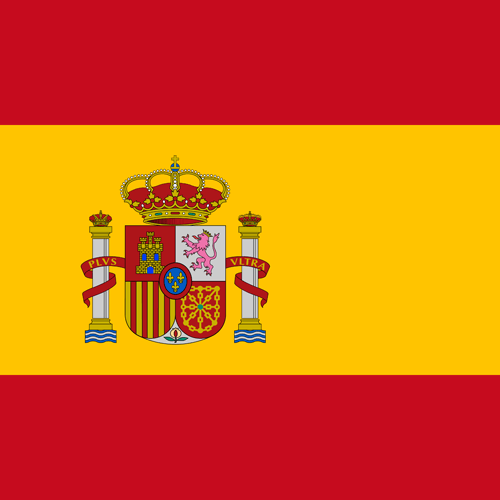}} & 7,010  & 98.5 \\
3  & Alexander Zverev & \raisebox{-0.3\height}{\includegraphics[width=0.9em]{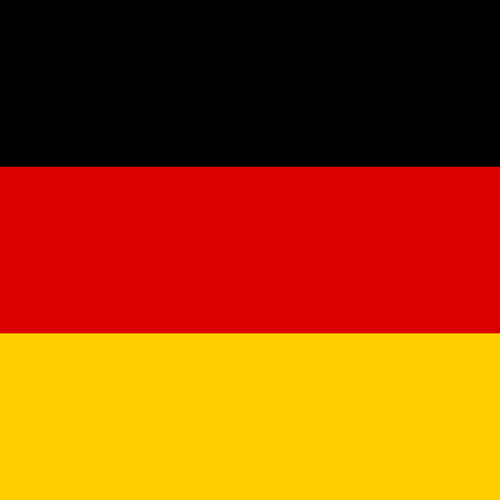}} & 7,915  & 97.6 \\
4  & Daniil Medvedev  & \raisebox{-0.3\height}{\includegraphics[width=0.9em]{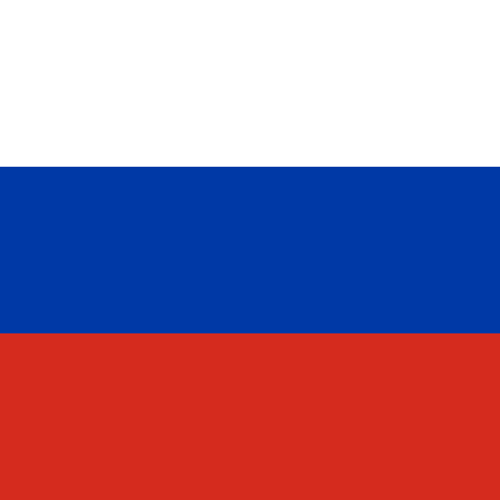}} & 5,030  & 97.1 \\
5  & Novak Djokovic   & \raisebox{-0.3\height}{\includegraphics[width=0.9em]{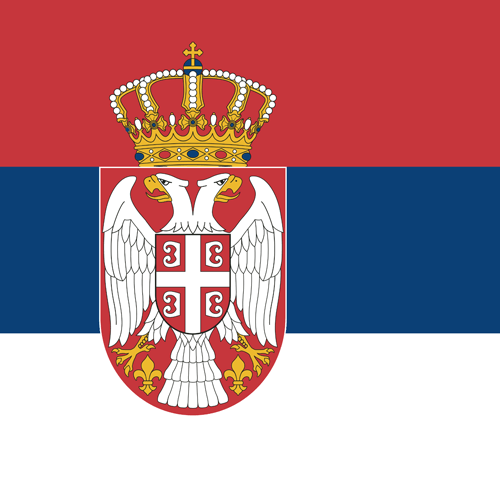}} & 3,910  & 96.7 \\
6  & Alex De Minaur   & \raisebox{-0.3\height}{\includegraphics[width=0.9em]{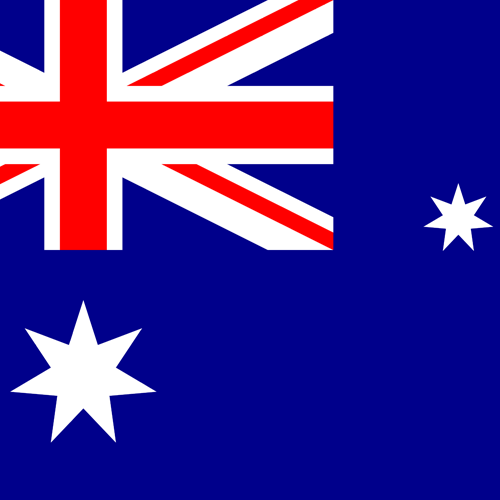}} & 3,745  & 96.2 \\
7  & Taylor Fritz     & \raisebox{-0.3\height}{\includegraphics[width=0.9em]{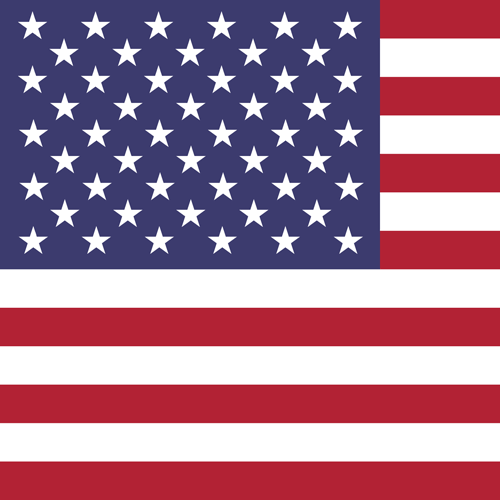}} & 5,100  & 95.2 \\
8  & Grigor Dimitrov  & \raisebox{-0.3\height}{\includegraphics[width=0.9em]{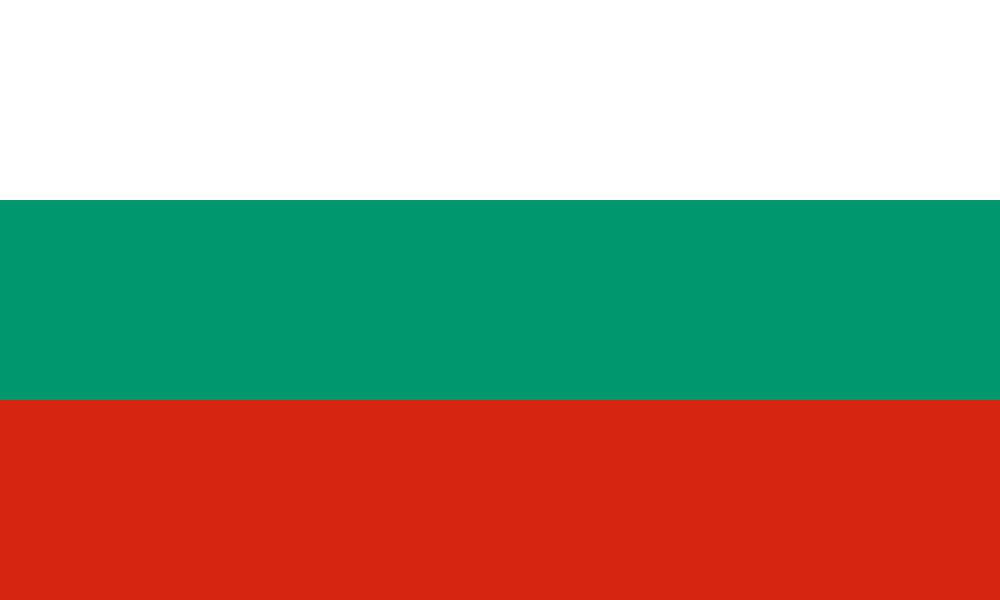}} & 3,350  & 94.8 \\
9  & Tommy Paul & \raisebox{-0.3\height}{\includegraphics[width=0.9em]{flags/usa.png}} & 3,145  & 93.8 \\
10 & Hubert Hurkacz   & \raisebox{-0.3\height}{\includegraphics[width=0.9em]{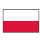}} & 2,640  & 93.3 \\
\hline
\end{tabular}
\caption{End of 2024 for the ATP. We rank the top $10$ by inferred posterior percentile and additionally report their ATP points.}
\label{tab:ATP_2024_rankings}
\end{table}

Summarizing players by single numbers is always going to be reductive and one should additionally consider uncertainty.
An obvious approach to uncertainty quantification is to report a variance-like measure.
Our analyses caution against this approach and show a considerably more nuanced picture.

A ``nice'' property of the Bradley-Terry model is that it is log-concave, hence, posterior distributions will be uni-modal.
This means we should be able to specify posteriors fairly accurately with a mean and a variance.
While this is indeed an algorithmically convenient fact---indeed, it is the key assumption of the expectation propagation algorithm of Ref.~\cite{trueskill}---it is additionally a very strong claim about what kind of uncertainty is possible.

Often players will have inconsistent records.
For example, suppose a player wins against several of the strongest players but then loses to some of the weakest.
There are two obvious possibilities: either this is a very strong player who had some bad luck, or, it is a weaker player that had good luck.
If the Bradley-Terry (or any log-concave) model is true then this is an impossible conclusion since posteriors must be uni-modal.
However, when we use an inferred kernel $b(x,y)$ such inferences are indeed possible.

\begin{figure}
    \centering
    \includegraphics[width=1\linewidth]{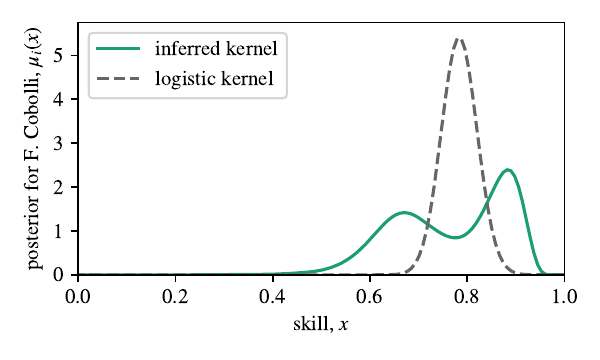}
    \caption{Inferred posterior distribution for the skill of Flavio Cobolli in year 2024. 
    We show the posterior distribution for both the Bradley-Terry model (i.e., logistic kernel) and the inferred kernel $b(x,y)$ found by the Chebyshev method, when fit to the data from 2021--2022.
    }
    \label{fig:cobolli}
\end{figure}

To see this in action, we look at the performance of Flavio Cobolli. 
Cobolli was a promising but somewhat inconsistent player during the 2024 season: he rose rapidly from outside the top 100 to finish the year inside the top 40 and reached his first ATP final in Washington.
Yet, his results across the year included both high-profile upsets and unexpected defeats.
In Fig.~\ref{fig:cobolli} we plot the posterior distribution for his skill, using both our inferred kernel $b(x,y)$ (from the Chebyshev method) and the logistic kernel of the Bradley-Terry model.
While both approaches roughly agree on the mean posterior skill, the representation of uncertainty is considerably changed.
The inferred kernel, with its multimodal posterior distribution, leads to a numerically superior fit to the observed game data.
Multimodality and nuanced uncertainty may be particularly important for emerging players who are developing and show variable performance.

\subsection{External validation against betting markets}
We believe our model is reasonable because it is a based on a principled Bayesian methodology.
However, sports provide a nice setting for testing these methods because we can compare with the odds offered by bookmakers; we test our model against the odds offered by Pinnacle (odds and match data were retrieved from \href{http://www.tennis-data.co.uk/alldata.php}{tennis-data.co.uk} \cite{tennis_data}).
Pinnacle has a reputation for tolerating winning players and integrating their feedback with internal models to achieve profitability despite the low margins \cite{pinnacle_ringer}, and for this reason their odds are less likely to be mispriced.

In order for our model to make predictions about future games one can compute the expected probability that player $i$ beats player $j$ given the inferred skill distributions
\begin{equation}
E[b(x_i, x_j)] = \iint \mu_{ij}(x, y)  b(x, y)  \mathrm{d}x \mathrm{d}y
\end{equation}
Note that for new players, i.e., those who have not participated in any observed games, the skill distribution will simply be the prior which is uniform.

We again set $b(x,y)$ to be the kernel inferred by the Chebyshev method on data from 2021--2022.
Then, for each day in 2023--2024 we make predictions about the outcomes of that day's matches by computing $E[b(x_i, x_j)]$ in the posterior induced by the previous $12$ months of matches.

We follow a simple betting strategy that compares model predictions against the odds offered by Pinnacle prior to the match.
A bet is placed if the expected profit is between $0$ and $100\%$.
If the model expects to make more than $100\%$ profit the bet is declined.
This is because the model only sees win–loss records and if model predictions are vastly different to the bookmakers, we assume that additional context is present, such as an injury.

\begin{figure}
    \centering
    \includegraphics[width=1\linewidth]{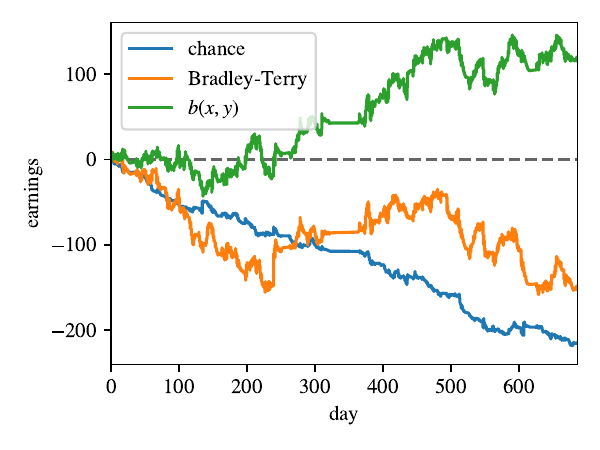}
    \caption{
    Simulated earnings over time with 3 different betting strategies, based only on the win-loss records.
    First, a ``chance'' betting strategy which, by definition, earns (negative) returns at the rate of the bookmaker's margins.
    Second, a strategy based on the Bradley-Terry model predictions. 
    Third, a strategy using the inferred kernel by the Chebyshev method.
    For the Bradley-Terry and kernel approach, bets are placed only if the model predicts between $0$ and $100\%$ profit.
    }
    \label{fig:betting_odds}
\end{figure}

In Fig.~\ref{fig:betting_odds} we see the outcome of three simulated betting strategies, each based only on the win-loss records of players over the preceding $12$ months.
Each strategy makes a fixed size bet on each game, and this size is set so that all strategies risk the same stake over the 2 year period.

The first strategy is to bet randomly on one of the two players. 
This leads to a fairly consistent loss corresponding to the bookmaker's margin.

The second strategy makes predictions using the Bradley-Terry model.
In the time period considered the Bradley-Terry model slightly outperforms random guessing.
Because this model is widely known and widely used, this is consistent with some level of conscious mispricing by the bookmaker.

Finally, using the inferred Chebyshev kernel to make predictions yields a significant improvement.
In fact, not only does this erase the margins but is actually profitable in the tested time period.
While we cannot know the ``true'' beliefs of the bookmakers, the fact that the inferred kernel approach is profitable suggests our predictions---based only on the win-loss records---are very close or even outperforming more sophisticated models.

\section{Discussion}

We have shown that, based only on the win-loss records, i.e., based only on the matrix
\begin{equation}
  w_{ij} = \text{number of times }i \text{ beat } j,
\end{equation}
we are able to infer a function that determines the probability that one player beats another, even in the sparse regime.
We present an efficient EM algorithm that achieves this and, additionally, returns the posterior distribution for each individuals strength percentile.

Our experiments on synthetic data demonstrate that both the Chebyshev and neural network approaches converge to kernel functions that closely match the ground truth.
While we cannot access ``ground truth'', experiments on real data show consistency between both approaches, which provides empirical evidence that the inferred kernels capture genuine underlying signals rather than artifacts of the method.
Finally, the experiment against bookmakers odds for men's professional tennis indicate the approach has good predictive accuracy.

Our approach is data driven and we place relatively weak constraints on the kernel functions.
Despite this, in Fig.~\ref{fig:real_data_kernels} we see a split between dominance hierarchies, which have steep almost step-like kernels, and competitive games, which have flatter kernels and hence larger upset probabilities.
We also see signs of location dependence: the inferred kernels are not constant along lines of constant $(y-x)$ so that, contra most work, assuming the kernel function can be written $b(x,y)=b(y-x)$ is not supported by the data.

By allowing for more flexibility in the model, we are able to represent more sophisticated kinds of uncertainty, such as the multi-modal uncertainty between a potentially strong-but-unlucky player or a weak-but-lucky one.
While this form of uncertainty surely seems worth considering,
standard models such as the Bradley-Terry mathematically forbid it---any log-concave distribution must have a single mode.

We have considered the most basic setting for ranking from pairwise comparisons.
There is a long body of work extending the simple ranking models into more sophisticated cases such as those with more possible outcomes, home-advantages, multiple players, multiple dimensions of skill, and so forth.
We anticipate no reason that our framework could not also be extended to these cases.

{\bf Code availability.} 
C\nolinebreak\hspace{-.05em}\raisebox{.4ex}{\tiny\bf +}\nolinebreak\hspace{-.10em}\raisebox{.4ex}{\tiny\bf +}
and Python code that implements our methods is available at \href{https://github.com/gcant/pairwise-comparison-inference}{https://github.com/gcant/pairwise-comparison-inference}


\appendix
\begin{table*}
\centering
\small
\begin{tabular}{cc}
\hline
\textbf{data set} & \textbf{description} \\
\hline
\multicolumn{1}{c}{\textbf{professional team sports}} \\
Basketball \cite{nba_data}   & National Basketball Association games 2015--2022 \\
Soccer     \cite{football_data}   & men's international association football matches 2010--2019 \\
\hline
\multicolumn{1}{c}{\textbf{professional individual sports}} \\
Tennis       \cite{tennis_data} & ATP men's singles games 2021--2022 \\
Online Chess \cite{chess_gm_data} & Chess.com games between GMs 2024 \\
\hline
\multicolumn{1}{c}{\textbf{amateur individual sports}} \\
Chess        \cite{lichess_data} & online chess games for players of all levels on lichess.com in 2016 \\
\hline
\multicolumn{1}{c}{\textbf{human}} \\
CS departments       \cite{clauset_inequality_2015} & doctoral graduates of one department hired as faculty in another \\
Business departments \cite{clauset_inequality_2015} & doctoral graduates of one department hired as faculty in another \\
\hline
\multicolumn{1}{c}{\textbf{animal}} \\
monkeys      \cite{vilette_comparing_2020}  & dominance interactions among a group of wild vervet monkeys \\
dogs         \cite{silk_dogs_2019}  & aggressive behaviors in a group of domestic dogs \\
baboons      \cite{franz_baboons_2015}  & dominance interactions among a group of captive baboons \\
mice         \cite{williamson_mouse_2016}  & dominance interactions among mice in captivity \\
\hline
\end{tabular}
\caption{Datasets used in the investigation.
         Several of these were used previously by Jerdee and Newman~\cite{jerdee_luck_2023}.}
\label{tab:dominance_dataset_descriptions}
\end{table*}

\end{document}